# Flux focusing with a superconducting nano-needle for scanning SQUID susceptometry

B. K. Xiang[1], S. Y. Wang[1], Y. F. Wang[1], J. J. Zhu[1], H. T. Xu[1], Y. H. Wang[1,2*]

*1. State Key Laboratory of Surface Physics and Department of Physics, Fudan University, Shanghai 200433, China*
*2. Shanghai Research Center for Quantum Sciences, Shanghai 201315, China*

\* To whom correspondence and requests for materials should be addressed. Email: wangyhv@fudan.edu.cn

**Abstract**

**Nano-fabricated superconducting quantum interference device (nano-SQUID) is a direct and sensitive flux probe useful for magnetic imaging of quantum materials and mesoscopic devices. Enabled by functionalities of superconductive integrated circuits, nano-SQUID fabricated on a chip is particularly versatile but spatial resolution has been limited by its planar geometry. Here, we use femtosecond-laser 3-dimensional (3D) lithography and print a needle onto a nano-SQUID susceptometer to overcome the limit of a plane-structure. The nano-needle coated with a superconducting shell focuses the flux both from the field coil and the sample. We perform scanning imaging using such a needle-on-SQUID (NoS) device on superconducting test patterns with topographic feedback. The NoS shows improved spatial resolution in both magnetometry and susceptometry over its planarized counterpart. This work serves as a proof-of-principle for the integration and inductive coupling between superconducting 3D nano-structures and on-chip Josephson nano-devices.**

**Introduction**

Superconducting quantum interference devices (SQUID) are one of the most sensitive magnetic detectors [1–4]. Nano-SQUIDs with miniaturized SQUID loop or pickup coil can be placed in close proximity with sample to enhance magnetic field sensitivity as well as to perform scanning microscopy [5–7]. This is important for the study of samples with small volume, especially two-dimensional quantum materials and quantum devices fabricated from these materials [8]. While the direct flux sensitivity of nano-SQUID is useful for imaging physical quantities like magnetization of a ferromagnet, vortex in a superconductor [9,10] or edge current in a quantum spin Hall insulator [11], susceptometry [12–15] is indispensable for probing spin correlations [16] or superfluid density [17–19], which are not observable from magnetometry.

Susceptometry performed by nano-SQUID utilizes the local magnetic field generated by a small field coil wound around the pickup loop to excite the sample [5]. The response to the excitation from the sample is proportional to its susceptibility and is detected through the pickup loop. Typically, a gradiometric geometry of the SQUID loop is necessary to suppress the mutual inductance between the field coil and the pickup loop [20]. Furthermore, a modulation coil is desired in order to linearize the flux signal by flux-locked-loop [21]. These requirements are only practical using a multi-layered nano-fabrication process on a planar substrate.

Although susceptometry is important for investigating 2D spin systems and superconductors, the on-chip nature of a nano-SQUID susceptometer has a serious limitation on its spatial resolution. The geometric constrain of a planar structure like a nano-SQUID chip prevents close proximity of the pickup and field coils to the sample [22,23]. It is also difficult to use the edge of a chip for height feedback, which are fundamental to the nanoscale spatial resolution in tip-based scanning probe microscopies [24]. A new design is needed to transcend a planar superconducting circuit without compromising the functionality of a nano-SQUID gradiometric susceptometer.

In this work, we demonstrate the successful integration of a 3D superconducting nano-needle with a nano-SQUID susceptometer on a chip. We report the design, fabrication, characterization and test imaging with such a needle-on-SQUID (NoS) probe. We show that the superconducting needle acts as a nano-flux-lens to focus the otherwise spreading flux lines both from the field coil and from the sample. The needle allows us to perform nanoscale topographic and magnetic imaging simultaneously. The spatial resolutions of the NoS magnetometry and susceptometry are both superior to those of the nano-SQUID susceptometer without the needle.

**Working principle of NoS and simulations**

The flux-focusing needle we study here is a cone-like structure with a thin superconducting shell that has opening at the apex and a slit on the sidewall (Fig. 1a). The body of the needle is non-magnetic and is situated directly on top of the front pickup loop of a nano-SQUID susceptometer. The superconducting shell is the main flux-focusing medium because Meissner screening of a superconductor prevents flux lines from going through the shell. The hole at the apex can be drilled to a much smaller size than that of the pickup loop so that only flux lines along the needle's axis ($z$) directly underneath the apex enter the needle. The sharp apex allows highly sensitive force microscopy so that the opening can be placed within 10 nm from sample under study. These qualities of the superconducting needle are useful for enhancing the spatial resolution in magnetometry. However, if there is no slit on the sidewall, Meissner screening current can flow around the hole at the apex and a small static magnetic field outside the needle will be completely expelled. To avoid that, a slit on the sidewall breaks the shell with a hole into a singly-connected superconducting sheet [25]. The slit should be made as narrow as possible to minimize flux leaking from it.

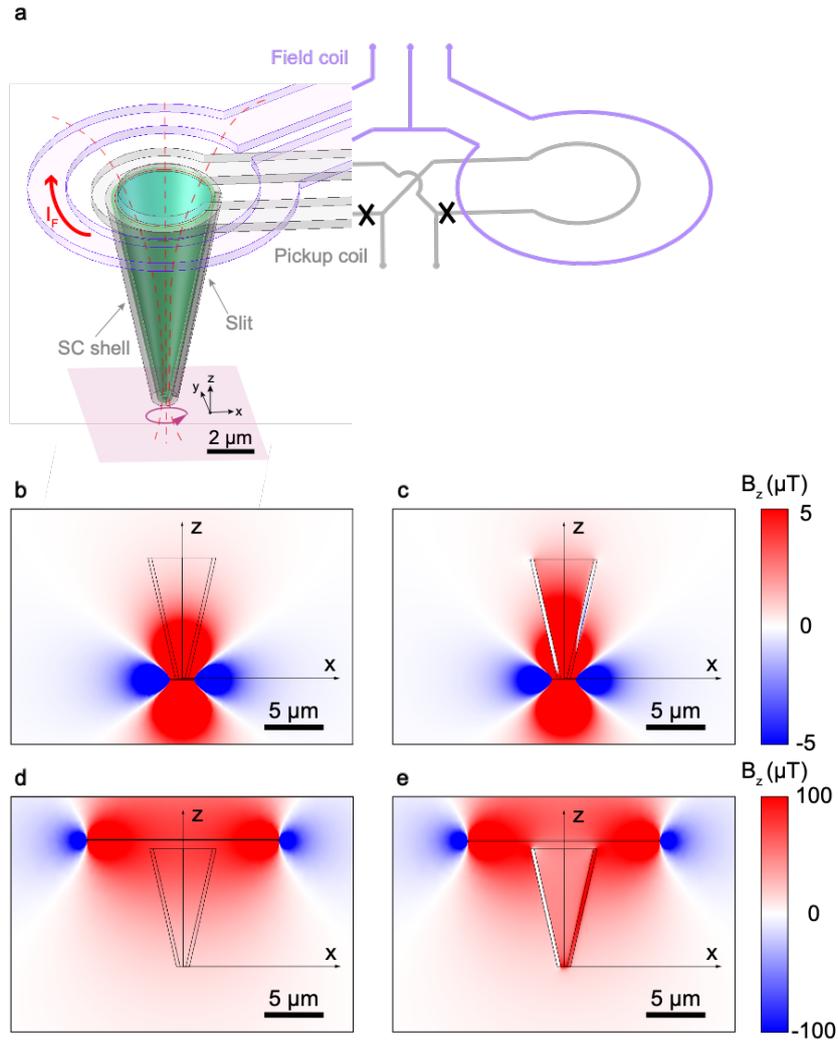

**Figure 1 Flux focusing on the nano-scale.** (a) Schematic diagram of the needle-on SQUID (NoS) device. The needle (green) is a dielectric with no effect on magnetic field distribution. It is covered with a superconducting (SC) shell (gray) with a hole at its apex and a slit on the sidewall, which breaks the shell into a singly-connected topology (see text). The front pickup coil (gray) of a gradiometric SQUID loop is at the base of the needle. It collects the magnetic flux collimated into the needle to be read out by the SQUID. Field coil (purple) generates an oscillating magnetic field focused through the needle onto the sample for susceptometry. (b and c) Finite element simulation of magnetic field in the $z$ direction ($B_z$) when there is a 1-μm-radius ring with current of 1 mA placed right beneath the hole (500-nm diameter) of the needle (10-μm height) with non-SC and SC shell, respectively. (d and e) Simulated $B_z$ distribution when applying a current $I_F =$ 1 mA in the field coil (16-μm diameter) with non-SC and SC shell, respectively. The asymmetry in the field distribution with the SC shell is caused by the slit on the sidewall.

In order to quantitatively examine the flux focusing effect from the superconducting nano-needle, we first perform finite element simulation of magnetic field distribution. We choose a needle height of 10 μm (sufficient for force microscopy) and a base diameter of 5 μm to match the pickup coil of our nano-SQUIDs. The superconducting shell is simulated by a medium with very small relative magnetic permeability (0.01). Shell thickness is set to 300 nm; the hole diameter at the apex is 500 nm and the slit width is 100 nm. The magnetic field distribution from a current ring (1 μm radius) placed under the apex of the needle is expectedly similar to a magnetic dipole field if the needle is non-superconducting (Fig. 1b). But when the shell is superconducting, $z$ component of the magnetic field ($B_z$) becomes much stronger inside the pickup loop (Fig. 1c and Fig. S1a). This is a strong indication that the nano-needle collimates part of the spreading flux lines generated by the current ring into the pickup loop.

The superconducting needle acts as a collimation lens not only for the flux from the sample but also for the flux generated by the field coil. This can be seen from the $B_z$ generated by flowing current through the field coil (Figs. 1d and e). The $B_z$ at the apex of the needle with a superconducting shell is much larger than that with a non-superconducting one (Fig. 1e and Fig. S1b). The asymmetry in $B_z$ is caused by the slit on the sidewall which leaks some flux outside the needle. For the same excitation intensity on the sample, the field applied through a superconducting shell impacts a much smaller sample volume which locates right under the opening of the needle for efficient detection. This property is potentially useful for scanning magnetic resonance imaging [26] using NoS.

**Nano-fabrication of NoS**

We describe the realization of NoS with nano-fabrication (Fig. 2). The main structure of the nano-needle is 3D-printed on the nano-SQUID chip using a home-built direct-writing photolithography system with a femtosecond Ti:Sap laser (Fig. S2). Since it is based on two-photon photopolymerization, femtosecond laser photolithography can produce nano-structures exceeding the diffraction limit of the infrared light employed [27,28]. The high resolution is important for achieving a sharp apex on the needle. Tens of microns thick photosensitive resin (SU8) is spin-coated on a nano-SQUID chip to start with (Fig. 2a). The body of a needle with the desired geometry is printed voxel by voxel onto the front pickup loop. The needle is hardened after curing at elevated temperature, resulting in a Young's modulus larger than $SiO_2$. The unwanted resin without exposure to the laser is washed off by developer (Fig. 2b).

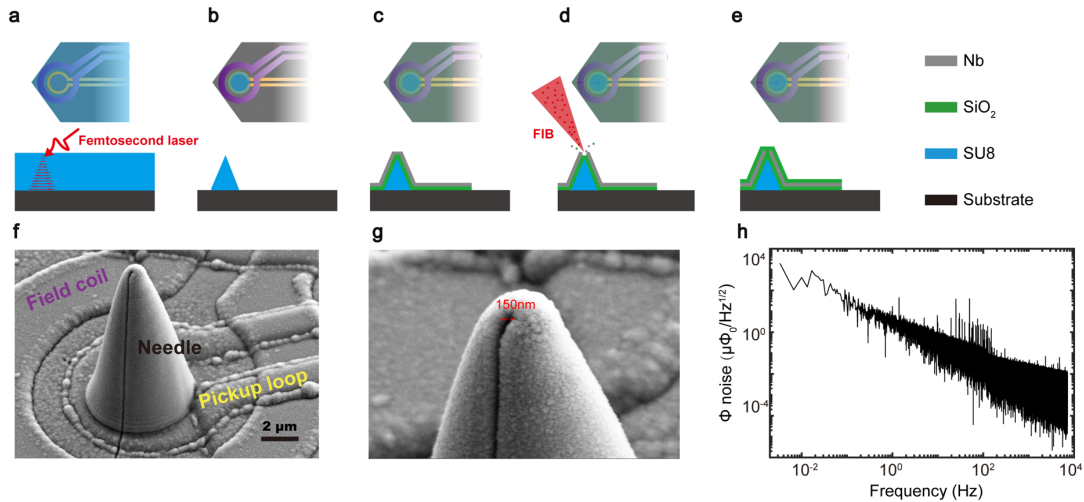

**Figure 2 Fabrication and characterization of NoS device.** (a) Printing 3D needle structure by femtosecond laser on a nano-SQUID chip. (b) Removing unexposed resin with a developer. (c) Deposition of 200 nm $SiO_2$ and 300 nm Nb as the superconducting shell. (d) Nano-machining of a hole on the apex and slit on the sidewall by focused ion beam (FIB). (e) Deposition of $SiO_2$ protection layer. (f, g) Scanning electron microscopy images of an exemplary NoS device. (h) Noise spectrum of an NoS obtained at 4.2 K under flux-locked-loop.

After constructing the main needle structure, we make the superconducting shell on top of it. We first deposit 200 nm $SiO_2$ as an insulation layer for the SQUID. Then we deposit 300 nm Nb by magnetron sputtering as the superconducting shell (Fig. 2c). Using focused ion beam (FIB) machining, we sculpture a hole at the apex of the needle and a slit on the sidewall (Fig. 2d). Finally, we deposit another 100 nm of $SiO_2$ to protect the shell from possible mechanical damage during scanning microscopy (Fig. 2e). Scanning electron microscopy images show a fabricated NoS with a sharp apex (Fig. 2f). The hole size of this particular device is 150 nm and the width of the slit is about 100 nm (Fig. 2g). After finishing the device fabrication, we characterize the noise performance of NoS. A successful device exhibits similar flux noise characteristics as that before the fabrication process (Fig. 2h).

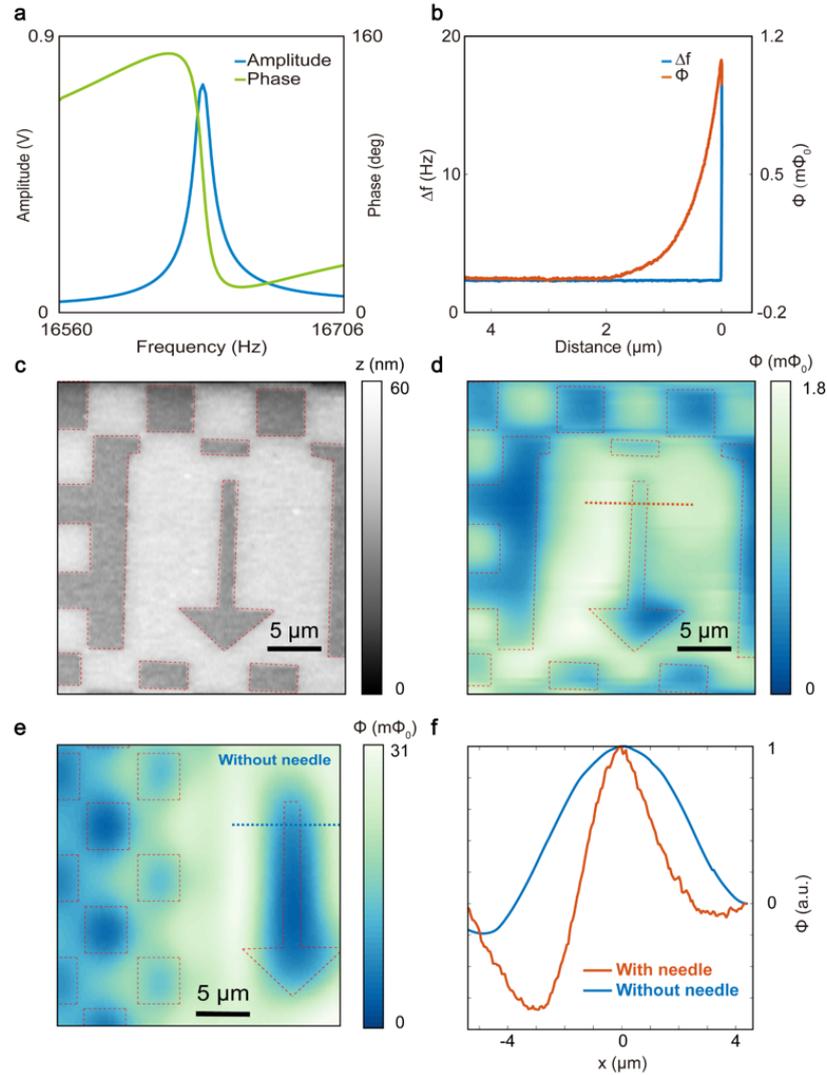

**Figure 3 Flux images of a Nb test sample.** (a) Amplitude and phase of the tuning fork-NoS assembly as a function of the drive frequency. (b) Approach curve of resonance frequency shift ($\Delta f$) and magnetic flux ($\Phi$) from NoS on a Nb test pattern. The NoS has an 800 nm opening on the needle and 5 μm pickup coil on the nano-SQUID. (c) AFM topography of the test sample measured by the same NoS. The image is obtained with height feedback to maintain a constant distance of several nanometers between the needle and the sample. The light regions are covered with 60 nm of Nb and the dark regions are film-free. (d) $\Phi$ image obtained simultaneously with (c). (e) Magnetic flux image on the same sample at a different area measured by a bare nano-SQUID susceptometer of 2 μm pickup coil without needle. The image is obtained at a constant scanning height of 500 nm. The arrow feature is the same as the one in (c). (f) Linecuts through the arrow patterns (straight dashed lines). $\Phi$ obtained by NoS is sharper, demonstrating the capability of flux focusing of the needle with a superconducting shell.

### Scanning imaging with NoS

We demonstrate the imaging capability of NoS on superconducting test samples. We attach an NoS to a quartz tuning fork (Fig. S3a) for atomic force microscopy (AFM) with the qPlus technique [29,30]. Resonant curve shows that the attachment of NoS shifts the resonance frequency of the tuning fork to $f_0$ around 16.6 KHz and reduces the overall quality factor (Fig. 3a). But it does not affect phase-locked loop operation of the NoS assembly for the height approach (Fig. 3b) and AFM imaging (Fig. 3c). We obtain flux signal in a flux-locked-loop together with the frequency shift during the sample approach (Fig. 3d). We use the demodulated flux signal at $f_0$ as the magnetometry signal ($\Phi$). The rise of $\Phi$ starts around 2 μm away from the sample and peaks where the needle touches the sample surface.

We set the tuning fork frequency to 1 Hz above $f_0$ for frequency modulated AFM after the approach (Fig. 3c). The height difference between the light regions with Nb film and the dark film-free region is 60 nm, consistent with the film thickness. Simultaneous magnetometry image (Fig. 3d) under the earth magnetic field 40 μT shows magnetic contrast consistent with the topography. The region covered with Nb shows higher signal than the film-free area (defined as zero flux) due to diamagnetic shielding by the film. As a direct comparison, we perform scanning imaging with a nano-SQUID susceptometer chip without needle on the same test sample (Fig. 3e). The scanned area of the sample is different from that in Figure 3d but the arrow pattern is the same size. Due to the lack of height feedback without the needle, we fix the scanning height at 500 nm from the touch-down point. The pickup loop on this nano-SQUID is 2 μm in diameter instead of the 5 μm used in the NoS. Furthermore, the 10-μm-high needle places the pickup loop of the NoS much farther away from the sample. And yet, we obtain a much sharper image by the NoS than by the nano-SQUID alone. For example, the arrow feature in magnetometry by NoS appears to be mostly within the bounds defined by topography. A line cut through the stem of the arrow shows a sharper variation (Fig. 3f). These magnetometry images strongly suggest that the superconducting needle on NoS is effective in focusing flux from the sample into the pickup loop.

We now present susceptometry imaging using a different NoS on a sample of superconducting square array. The nano-SQUID and the needle have the same parameters as the first one, except for a 500 nm opening on the needle and a functioning field coil 16 μm in average diameter. We flow 0.5 mA alternate current at 1126 Hz through the field coil and demodulate the $\Phi$ signal at this modulation frequency for susceptibility. The topographic image shows slightly rounded corners on the square and lower signal-to-noise (Fig. 4a) than the topographic image obtained by the first NoS (Fig. 3a). This suggests that the needle is slightly blunt after extensive scanning. The Nb squares in the magnetometry (Fig. 4b) and the susceptometry (Fig. 4c) images appear even more circular and smaller than the actual boundary of the squares. Rounded corners were also seen in a previous work imaging a checker-board square pattern with the same Nb thickness using a nano-SQUID susceptometer on a chip [14].

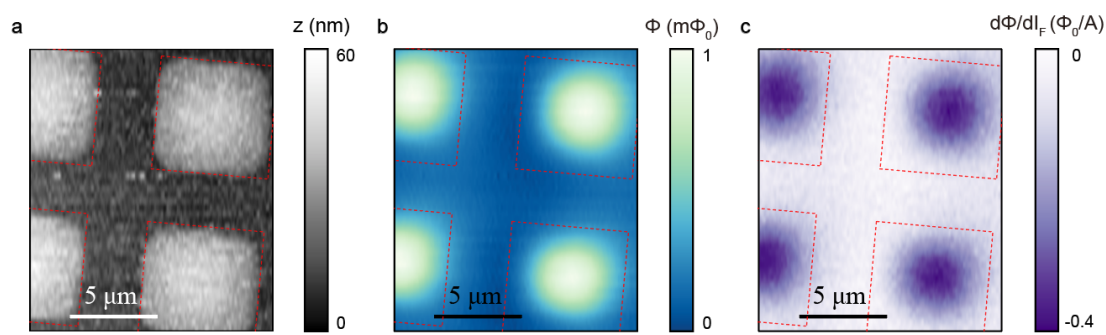

**Figure 4 Susceptometry of Nb square array.** (a) AFM topography of a 5 μm Nb square array obtained by a NoS with a 5-μm-diameter pickup coil and 16-μm-diameter field coil. The opening on the needle is 500 nm. Light squares are covered with 60 nm of Nb. The red-dashed lines are the boundaries of the squares. (b) Magnetic flux image of the squares. (c) Susceptometry image obtained simultaneously by applying 0.5 mA current at 1126 Hz through the field coil. Squares with the Nb film exhibits diamagnetic susceptibility.

In that work, the susceptometry images were distorted when 150-nm-diameter pickup loop was used with a 2-μm inner-diameter field coil. The large disparity between the sizes of the pickup loop and the field coil may have contributed to susceptometry artifacts when the square array has a similar size as the field coil. Noting that the opening at the needle which collects the flux is more than 10 times smaller than the field coil, we may expect the artifact is also present in the susceptometry by NoS. And yet, such an artifact is clearly absent (Fig. 4c), suggesting that the size disparity between flux collection and excitation is avoided. This fact implies that the detected susceptibility is mainly induced by a localized magnetic field which the superconducting needle is responsible for creating.

We discuss several possibilities for the origin of rounded corners in the magnetic images. First, the characteristic magnetic length scale of superconducting thin film plays a large part. It is described by the Pearl length $\Lambda = 2\lambda^2/d$ [31], where $\lambda$ is London penetration depth and $d = 60$ nm is film thickness. For Nb $\lambda = 140$ nm at 1.8 K [32], we get $\Lambda = 660$ nm, which is larger than the opening of the needle. This length scale determines the width of the Meissner screening current, which is ultimately responsible for the magnetic signal. In addition, the Meissner current does not flow against the edge of the square. At the corners, it shifts further away from the edge to reduce its free energy. Therefore, the magnetic images appear circular than squared. Since the susceptibility also comes from (locally) exciting the Meissner current, a large $\Lambda$ can also smear out sharp geometric features even if the superfluid density is uniform across the square.

Geometric effect at the apex of the needle may contribute to broadened magnetic features as well. We estimate from the scanning electron microscopy image that the radius of curvature at the apex is about 500 nm after fabrication. The apex may become blunter after extensive usage (Fig. 4a). This can mechanically shift the opening from the apex, which allows in-plane magnetic flux to enter. Such misalignment may affect the magnetic signal when topography changes quickly, such as at a step edge. Lastly, since the flux generated by the field coil is focused through the apex of the needle (Fig. 1e), the magnetic field there may exceed the lower critical field of the superconducting shell. This results in reduced shielding and effectively increases the area of the opening, decreasing the overall resolution both in magnetometry and susceptometry.

In principle, the physical resolution limit of using a nano-needle is determined by the penetration depth of the superconducting shell. Our current work is clearly far away from such a limit set by the penetration depth of Nb. Magnetic behavior of the test sample and geometric effect of the needle as discussed above need to be experimentally addressed. We leave the optimization of the nano-needle to further enhance the spatial resolution in susceptometry for future works.

**Conclusion**

In conclusion, we successfully integrate superconducting needle on nano-SQUID susceptometers by femtosecond laser 3D nanolithography. We demonstrate the flux-focusing capability of such NoS devices. By performing topography, magnetometry and susceptometry imaging of superconducting patterns, we show that the NoS's are superior over nano-SQUID susceptometers. Our method of constructing 3D superconducting structures on planar superconducting circuits may find broad applications in flux-based mesoscopic quantum devices.

**Acknowledgement**

We would like to acknowledge support by National Natural Science Foundation of China (Grant No. 11827805 and 12150003), National Key R&D Program of China (Grant No. 2021YFA1400100 and 2017YFA0303000) and Shanghai Municipal Science and Technology Major Project (Grant No. 2019SHZDZX01). All the authors are grateful for the experimental assistance by Y. P. Pan, Y. Feng, X. D. Zhou and W. X. Tang.

**Data availability**

The data that support the findings of this work are available from the corresponding authors upon reasonable request.

*Magnetic Penetration Depth on the Thickness of Superconducting Nb Thin Films*, Phys. Rev. B **72**, 064503 (2005).